\documentclass[comsoc, journal]{IEEEtran}
\usepackage[utf8]{inputenc}
\usepackage[T1]{fontenc}

\usepackage{subcaption}
\usepackage{caption}
\usepackage{graphicx}
\usepackage{float}
\usepackage{hyperref}

\usepackage{tabularx}

\usepackage[noadjust]{cite}

\usepackage{array, makecell, multirow, booktabs}
\newcolumntype{P}[1]{>{\centering\arraybackslash}p{#1}}
\setcellgapes{3pt}

\newcommand{\eg}{\textit{e.g.,~}}
\newcommand{\ie}{\textit{i.e.,~}}

\usepackage{color, xcolor}

\newcommand{\revision}{\color{black}}
\newcommand{\rone}{\color{black}}

\usepackage{color, soul}
\usepackage{xspace}

\IEEEpubid{\begin{minipage}{\textwidth}\ \\[32pt] \centering
  \footnote {This manuscript has been accepted for publication by the IEEE Wireless Communications Magazine © 2022 IEEE. Personal use of this material is permitted. Permission from IEEE must be obtained for all other uses, in any current or future media, including reprinting/republishing this material for advertising or promotional purposes, creating new collective works, for resale or redistribution to servers or lists, or reuse of any copyrighted component of this work in other works.}
\end{minipage}}

\begin{document}

% \title{Enabling Biomechanical Research Through Prominent Networking and Computing Technologies}

 \title{The Promise and Challenges of Computation Deduplication and Reuse at the Network Edge}

%\title{Networking and Computing in Biomechanical Research: Easier Said than Done}

\author{
	Md Washik Al Azad and Spyridon Mastorakis 
	
	\thanks{Md Washik Al Azad and Spyridon Mastorakis (corresponding author) are with the University of Nebraska at Omaha, USA.}
}

\markboth{}{}

\maketitle

\begin{abstract}

In {\rone edge computing deployments,} where devices may be in close proximity to each other, these devices may offload similar computational tasks (\ie tasks with similar input data for the same edge computing service or for services of the same nature). This results in the execution of duplicate (redundant) computation, which may become a pressing issue for {\rone future edge computing environments}, since such deployments are envisioned to consist of small-scale data-centers at the edge. To tackle this issue, in this paper, we highlight the importance of paradigms for the deduplication and reuse of computation at the network edge. Such paradigms have the potential to significantly reduce the completion times for offloaded tasks, accommodating more users, devices, and tasks with the same volume of deployed edge computing resources, however, they come with their own technical challenges. Finally, we present a multi-layer architecture to enable computation deduplication and reuse at the network edge and discuss open challenges and future research directions.
\end{abstract}

\IEEEpeerreviewmaketitle

\section{Introduction}
\label{sec:intro}

Edge computing has emerged as a paradigm to bring computing resources physically close to end-users in an effort to address the increasing needs of {\revision applications for} {\rone the} low-latency processing of data generated by user devices, {\revision such as mobile phones, Augmented Reality (AR) headsets, {\rone and} Internet of Things (IoT)~\cite{shi2016edge}}. Edge computing deployments are envisioned to consist of small-scale data-centers at the edge of the network~\cite{satyanarayanan2017emergence}. {\rone At the same time,} such deployments may target large-scale use-cases (\eg smart cities with hundreds of thousands or millions of residents). %As a result, mechanisms other than brute-force hardware advancements (\ie \enquote{throwing more hardware to the problem}) may be needed to accommodate such use-cases.
%
%brute-force hardware advancements (\ie \enquote{throwing more hardware to the problem}) may not  
%
In such use-cases, several devices may be in close proximity to each {\revision other, offloading tasks} for ``similar'' computation (\ie tasks with similar input data for the same edge computing service {\revision or services} of the same nature) to the edge~\cite{guo2018foggycache}. This can result in the execution of massive amounts of duplicate (redundant) computation, limiting the number {\rone of devices and tasks} that can be accommodated by edge computing deployments. Overall, we expect the execution of duplicate computation to become a pressing issue for future edge computing deployments given {\rone the expected small scales} of edge data-centers and the need to accommodate large-scale use-cases. 

In this paper, we highlight the promise of paradigms for the deduplication and reuse of computation at the network edge. In such paradigms, the results of previously executed tasks are stored {\rone with the goal to be reused and satisfy similar offloaded tasks, instead of executing similar tasks from scratch.} {\revision The process of deduplication provides the means to infer whether reuse is possible by determining whether tasks similar to the offloaded ones have been previously executed and stored at the edge.} As a result, such paradigms essentially ``trade'' storage for computing resources, having the potential to: (i) significantly reduce the completion time of offloaded tasks; and (ii) accommodate more users, devices, and tasks with the {\rone same volumes} of deployed edge computing resources. However, such paradigms come with their own technical challenges, which need to be addressed for their realization.

Our contribution in this paper is two-fold: (i) we highlight the promise that computation deduplication and reuse holds {\rone for edge computing environments} and the need for paradigms for deduplication and reuse that consider the distributed nature of {\revision such environments}, a key observation that prior research has overlooked; and (ii) we present a multi-layer architecture to enable the pervasive computation deduplication and reuse at the network edge along with promising proof-of-concept evaluation results. 

{\revision The rest of this paper is organized} as follows: in Section~\ref{sec:background}, we discuss the importance of computation deduplication and reuse for edge computing deployments. In Section~\ref{sec:design}, we present the design of a multi-layer architecture for computation deduplication and reuse. In Section~\ref{sec:discussion}, we present open challenges and future research directions, and, in Section~\ref{sec:conclusion}, we conclude our work.

\section{Why Computation Deduplication and Reuse at the Edge Are Important?}
\label{sec:background}

\begin{figure}[th]
    \centering
    \includegraphics[width=1\linewidth]{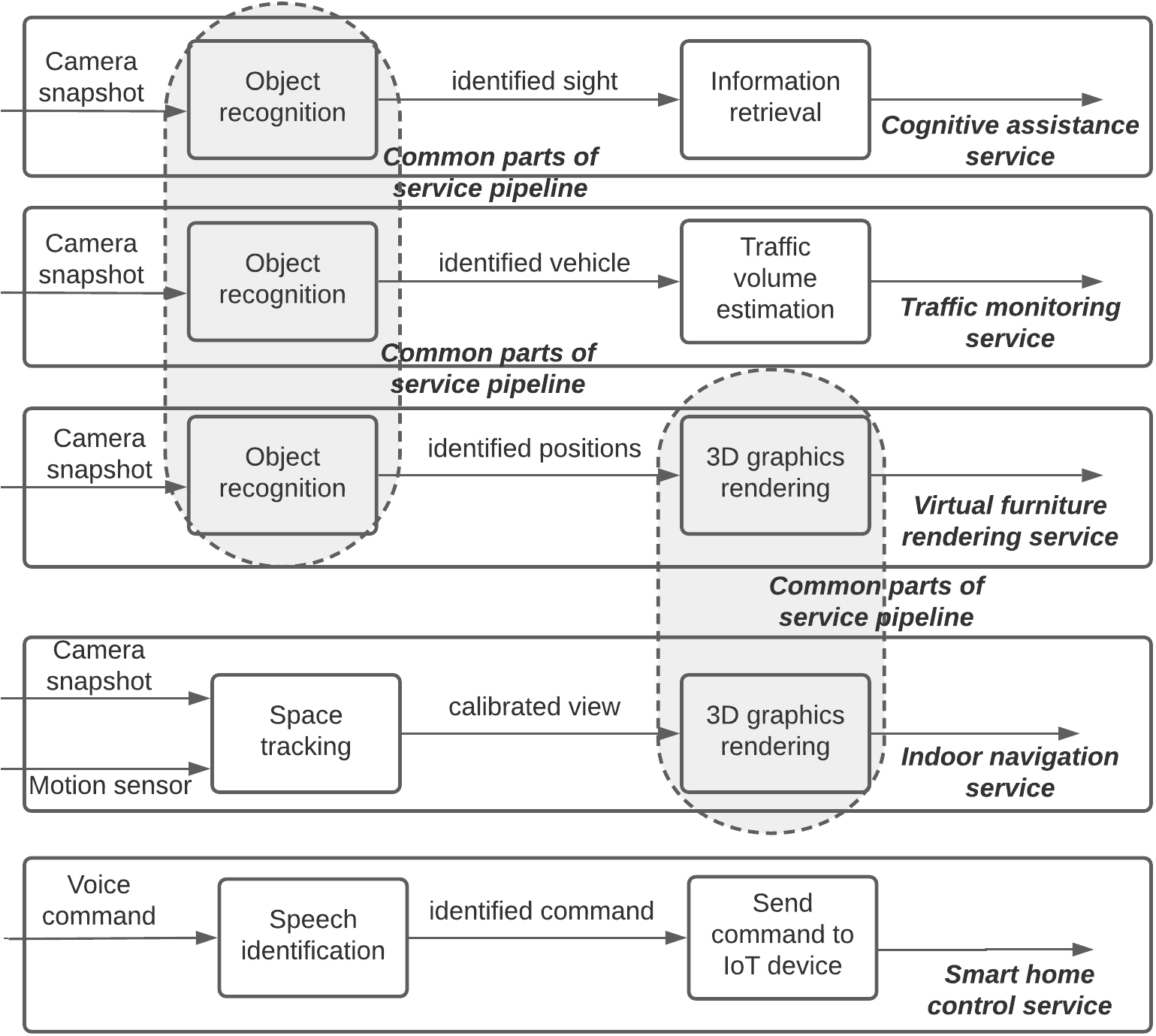}
    \caption{  Processing pipelines for four applications.}
    \label{fig:pipelines}
    %\vspace{-0.2cm}
\end{figure}

With the projected growth of the number of IoT, mobile, and other devices at the edge, several devices may be in close proximity to each other. {\rone In such environments, redundant computation may occur, since temporal, spatial, and semantic correlation may exist between the input data of offloaded tasks. Devices may request the execution of the same services/processing functions offered at the edge with similar data as the inputs of these services/functions. In addition, available edge services may have processing components in common. 
 
For example, a cognitive assistance application may be used on mobile phones or AR headsets to recognize the environment in the captured camera snapshots or AR scenes. In this context, visitors of famous sights all around the world may use this application to capture pictures/scenes of a sight with their mobile phones or AR headsets. These pictures/scenes} are offloaded to a nearby edge server where {\rone a cognitive assistance service (Figure~\ref{fig:pipelines}) identifies} the {\revision depicted sight} and returns information and content about the identified sight to visitors (\eg podcasts and videos related to the sight, the story behind the sight). {\rone In this scenario, visitors may request} the same computation with similar inputs (\eg pictures of the same sight from different angles or distances), {\rone thus resulting in the same output} (the received information and content about the same sight). {\rone At the same time, this edge service may share processing components (\eg object recognition as illustrated in Figure~\ref{fig:pipelines}) with other edge services. Such services may include: (i) a service that estimates the volume of vehicle traffic based on snapshots captured by Closed-Circuit TeleVision (CCTV) cameras. Since such cameras may capture multiple snapshots every second, consecutive snapshots (\eg snapshots $n-1$ and $n$ in Figure~\ref{fig:arch}) may be highly similar; and (ii) a service used by an application that renders virtual furniture at certain positions to visualize furnished spaces~\cite{guo2018potluck}. Subsequently, an application for indoor navigation may render a virtual map to help users navigate buildings or stores. As a result, an edge service to process camera snapshots in this context may have a 3D graphics rendering component in common with the virtual furniture rendering service. Finally, in smart homes equipped with IoT devices, users can control these devices through voice commands. In such cases, residents of the same or nearby homes may invoke semantically similar commands that result in the same action (\eg turning on the light in a room). To this end, the results of the corresponding edge service (Figure~\ref{fig:pipelines}) can be shared/reused among multiple users.}

%\begin{figure}[th]
%    \centering
%    \includegraphics[width=1.03\linewidth]{fig/CCTV.pdf}
%    \caption{\small A CCTV camera capturing snapshots of vehicle traffic. These snapshots are offloaded to the edge, so that the number of cars in each snapshot is detected and the volume of traffic is estimated.}
%    \label{fig:cctv}
%    \vspace{-0.2cm}
%\end{figure}

%{  Overall, redundant computation may occur in scenarios where temporal, spatial, and semantic correlation may exist between the input data of offloaded tasks. Additional examples  may include: (i) smart homes equipped with IoT devices, where users can control these devices through voice commands. In such cases, residents of the same or nearby homes may invoke semantically similar commands that result in the same action (\eg turning on the light in a room); and (ii) AR applications, which, for example, can be used by first responders wearing AR headsets during a disaster to identify survivors and other objects. In such cases, first responders working in groups may capture similar scenes (\eg scenes of the same office space in a building on fire from different angles). As a result, running an edge service for the detection of survivors and objects in such scenes may yield the same execution results.}

At the same time, given that next-generation applications may require {\revision ultra-low response times} (\eg AR may require response times less than 10ms), the deduplication and reuse of computation can speed up the {\revision execution of tasks offloaded by user devices,} since the execution results of previous similar tasks can be reused instead of executing {\rone computation-intensive tasks from scratch.} This also ensures that the available edge computing resources are effectively utilized given their potentially limited scale by essentially trading storage for computing resources (\ie to store and reuse previously executed tasks and their results). As a result, the reuse of computation has the potential to increase the capacity of edge computing deployments in terms of being able to accommodate more tasks, users, and devices with a fixed amount of physical edge computing resources. 

\subsection{Computation Deduplication and Reuse: Making the Network Part of the Solution} 
\label{sec:networking}

{\rone The nature of edge computing deployments is distributed in the sense} that they consist of several edge servers for fault tolerance, scalability, and load balancing purposes. As a result, each computing service at the edge (\eg object detection, face recognition) may be offered by multiple edge servers. In the scenario illustrated in Figure~\ref{fig:arch}, if consecutive snapshots captured by a CCTV camera are offloaded to different edge servers for processing (\eg snapshot $n-1$ is offloaded to server A, while snapshot $n$ is offloaded to server B), then computation deduplication and reuse will not be possible. 

In other words, in realistic edge computing deployments, where each computing service may be offered by multiple edge servers, it is vital that the {\rone edge network infrastructure can facilitate the reuse of computation.} To achieve that, the edge network infrastructure needs to forward computational tasks for the same computing service and with similar input data to the same edge server. Essentially, the network needs to identify and forward tasks with similar input data with minimal overhead and performance penalty, calling for solutions that expose data similarity {\rone semantics at the network layer in a light-weight manner.} %{\revision In the context of our CCTV scenario, background removal approaches could also be utilized to identify similar snapshots. %~\cite{zivkovic2004improved}.
%However, such approaches may be expensive, while they may also make the exposure of data similarity semantics to the network layer cumbersome.}

%The overhead on users and their (potentially resource-constrained) devices should be minimal as well, 

{\rone In Table~\ref{table:reuse}, we present representative studies that have explored computation deduplication and reuse in edge computing deployments.} {\revision Cachier~\cite{drolia2017cachier} proposed optimizations of edge server caches} leveraging the spatiotemporal locality {\revision of user requests} for computation. Potluck~\cite{guo2018potluck} proposed the deduplication and reuse of computation across different applications running on the same user device, while FoggyCache~\cite{guo2018foggycache} explored the reuse of computation at edge servers across different user devices. Coterie~\cite{meng2020coterie} exploited the similarity among background environment frames in multi-player Virtual Reality applications, so that headsets can cache and reuse similar frames. However, all these approaches did not consider the challenges stemming from the distributed nature of edge computing deployments. ICedge~\cite{mastorakis2020icedge} proposed a preliminary design to facilitate computation reuse with the assistance of the edge network infrastructure in distributed edge computing deployments. However, in ICedge, computation {\revision deduplication and reuse are unlikely} to happen directly in the network, while each application {\revision may expose different computation reuse semantics to the network}, making task forwarding complicated. {\rone To this end, there is a need for solutions that allow the edge network infrastructure to forward tasks from different applications based on the same semantics} while facilitating pervasive computation deduplication and reuse---at user devices, within the edge network infrastructure, and at edge servers. 

\begin{table*}[t]
\caption{Studies on computation deduplication and reuse in edge computing environments.}
\label{table:reuse}
\centering
%\resizebox{\columnwidth}{!}{
\begin{tabular}{|c|c|c|c|c|c|c|}
\hline
\textbf{Solution}                                                                     & \textcolor{black}{~\cite{drolia2017cachier}}
& \textcolor{black}{~\cite{guo2018potluck}}                                                                 & \textcolor{black}{~\cite{guo2018foggycache}}                                                           &
\textcolor{black}{~\cite{meng2020coterie}}                                                           & \textcolor{black}{~\cite{mastorakis2020icedge}} &
\textcolor{black}{Our approach}\\ \hline
\textbf{\begin{tabular}[c]{@{}c@{}}Are Deduplication\\ and Reuse Possible?\end{tabular}} 
& \begin{tabular}[c]{@{}c@{}}Partially--at\\edge servers\end{tabular}
& \begin{tabular}[c]{@{}c@{}}Yes--at user\\ devices\end{tabular}          &
\begin{tabular}[c]{@{}c@{}}Yes--at edge\\ servers\end{tabular}          &
\begin{tabular}[c]{@{}c@{}}Yes--at user \\ devices\end{tabular}  &     
\begin{tabular}[c]{@{}c@{}}Yes--at edge\\ servers\end{tabular} &
\textcolor{black}{\begin{tabular}[c]{@{}c@{}} Yes--at user devices, \\ edge network \\ infrastructure, and \\ edge servers\end{tabular}} \\ \hline
\textbf{\begin{tabular}[c]{@{}c@{}}Focus\end{tabular}}                     
& \begin{tabular}[c]{@{}c@{}}Optimizations of \\ edge servers' \\ caches\end{tabular} & \begin{tabular}[c]{@{}c@{}}Optimizations of \\ user devices' \\ caches\end{tabular} & \begin{tabular}[c]{@{}c@{}}Cross-device\\ deduplication\end{tabular} &
\begin{tabular}[c]{@{}c@{}}Exploit inter-frame \\ similarity in Virtual \\ Reality\end{tabular} &\begin{tabular}[c]{@{}c@{}}Network\\ architecture\end{tabular} 
& \textcolor{black}{\begin{tabular}[c]{@{}c@{}} Multi-layer \\ architecture for \\ deduplication \& reuse \end{tabular}} \\ \hline
\textbf{\begin{tabular}[c]{@{}c@{}}Considers Distributed \\ Edge Computing \\ Deployments?\end{tabular}}           & No                                                             & No                                                                & No    
& No   
& Partially
& \textcolor{black}{Yes} \\ \hline
\end{tabular}
\end{table*}

\subsection {Goals and Technical Challenges}
\label{sec:challenges}

The fundamental goal {\revision to be achieved by} solutions for pervasive reuse of computation at the edge is imposing minimal overheads on {\rone (potentially resource-constrained) user devices, the edge network infrastructure, and the edge servers} so that users and edge computing providers can receive the substantial benefits of computation reuse. These benefits feature improved response times and the ability to accommodate increased numbers of tasks, users, and devices with a fixed amount of physical edge computing resources. Based on our analysis in the prior parts of this section, we can conclude that solutions for computation deduplication and reuse at the edge need to tackle the following challenges:

\begin {itemize}

\item Impose minimal overheads on users, their devices, and the network for the identification and forwarding of tasks with similar input data to the same edge server for processing and reuse. {\revision At the same time, edge servers must efficiently search for similar previously executed tasks and execute incoming tasks only if previously executed tasks cannot be reused.}

\item Reuse previous {\rone tasks and their results accurately.} In other words, the execution of an incoming task $t$ and a reused task $t_{reused}$ {\revision with similar input data} must yield the same execution results, so that the operation {\revision of applications} and the user-perceived quality of experience are not negatively impacted. {\rone This also applies to cases of edge services that have processing components in common, so that the results of these common components (intermediate results) can be shared among multiple edge services (partial reuse). }

%{  We discuss extensions of this notion in Section~\ref{sec:discussion}.}

\end {itemize}

\section{A Multi-Layer Architecture for Computation Deduplication and Reuse} 
\label{sec:design}

\begin{figure}[th]
    \centering
    \includegraphics[width=1.05\linewidth]{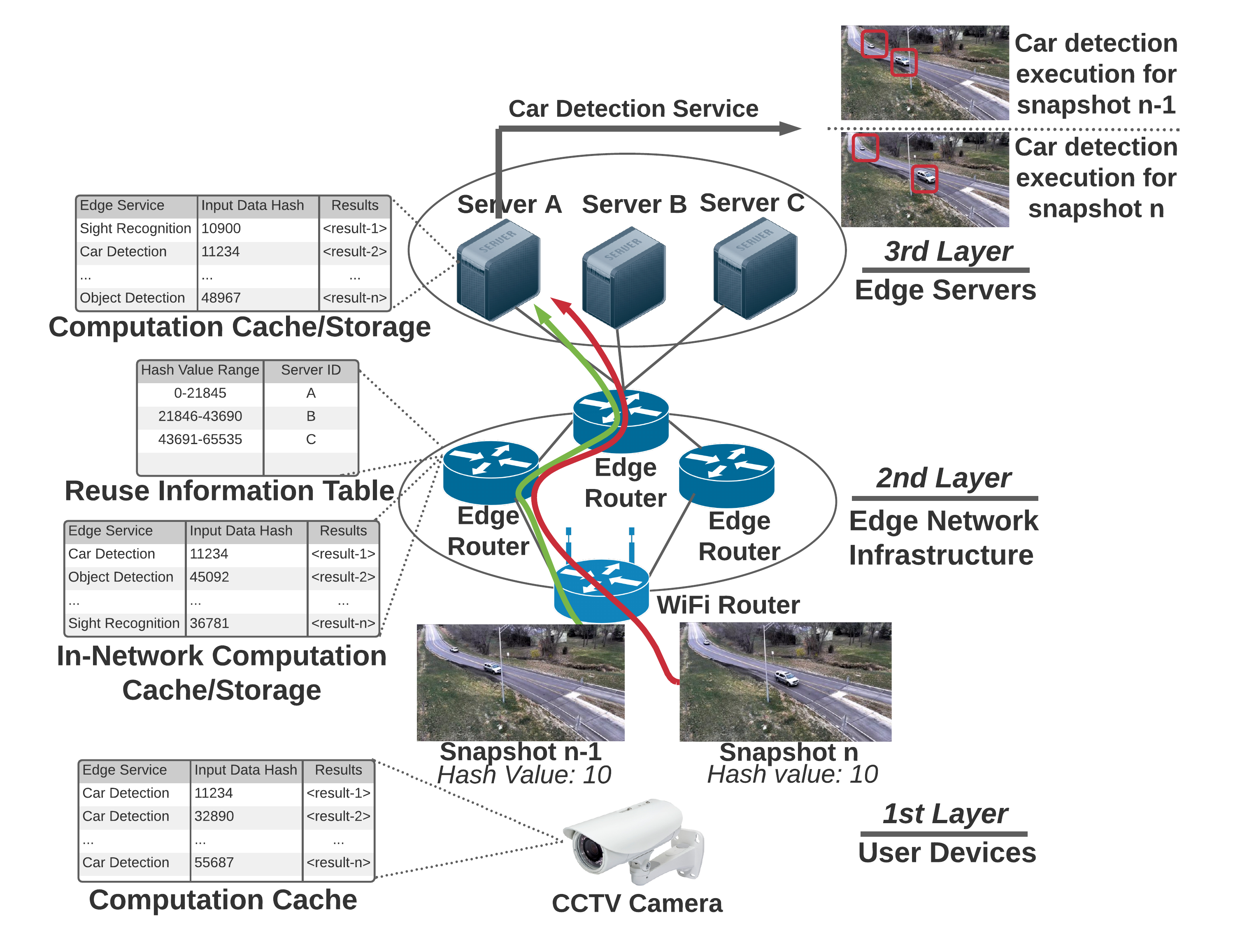}
    \caption{\small { A CCTV camera capturing snapshots of vehicle traffic, which are offloaded to the edge, so that the number of cars in each snapshot is detected and the volume of traffic is estimated. Consecutive snapshots may be similar, thus yielding the same output once processed through a car detection service at edge servers and resulting in the execution of duplicate computation. To enable pervasive computation deduplication and reuse at the network edge, we propose a multi-layer architectural design (Section~\ref{sec:design})}.}
    \label{fig:arch}
    %\vspace{-0.2cm}
\end{figure}

In this section, we present a hierarchical architecture for the deduplication and reuse of computation at the network edge (Figure~\ref{fig:arch}). {\rone The first layer of the architecture consists of user devices, which can cache the results of previously offloaded tasks depending on the availability and capacity of their resources}. The second layer consists of the edge network infrastructure, which identifies and forwards similar tasks to {\rone edge servers that can reuse previous computation, as well as features repositories for} caching previous tasks and their execution results directly in the network. The third layer consists of edge servers, {\revision which receive offloaded tasks, search for previously executed tasks to reuse,} and execute and store the results of incoming tasks when reuse is not possible. 

In the example of Figure~\ref{fig:arch}, snapshot $n-1$ captured by a CCTV camera is offloaded by the camera and is forwarded by the edge network infrastructure to edge server A for the detection of the number of vehicles in the snapshot. Snapshot $n$ captured by the CCTV camera will first have an opportunity to reuse the results of previous tasks stored in the computation cache of the CCTV camera itself {\revision (if the camera has adequate resources to do so)}. If this is not possible, the snapshot is offloaded to the edge, where it can reuse the results of previous tasks from in-network computation caches/storage. If this is not possible, the snapshot will be forwarded by the edge network infrastructure to the same edge server as snapshot $n-1$ (server A), where the execution results of the processing of snapshot $n-1$ may be reused.

%The edge servers may be one hop away (\eg directly attached to base stations) or a few hops away from user devices (\eg at the edge of the core network).

\subsection {Layer 1: User Devices}

The first layer of our architecture consists of user devices. Each device may run one or more applications and offload computational tasks to edge servers. {\rone Before offloading a task, each device needs to create a notion of how similar the task may be compared to previous tasks,} essentially aiding all the layers of our architecture to find previous tasks that can be reused in a light-weight manner. This can happen through fast and space-efficient mechanisms, such as Locality Sensitive Hashing (LSH) and Feature Hashing (FH). LSH is a technique that allows to search for the nearest neighbor of a data item $i$ by applying a hash function $h$ to $i$ and using the resulting hash $h(i)$ as the index of a bucket of a hash table to be searched for the nearest neighbor(s) of $i$. Furthermore, FH enables the vectorization of features extracted from data items (by applying a hash function $h$ to the extracted features), {\rone while the resulting vector can be further hashed through LSH to cluster data items with similar feature values.}

%{\revision More specifically, LSH is an algorithmic technique that hashes similar data items into the same buckets of a hash table, so that data clustering and nearest neighbor search operations are enabled. In the same manner, } 

Through the application of such hashing techniques on the input data {\revision (\eg images, videos, voice commands)} of tasks: (i) tasks with similar input data will likely be assigned the same hash value; and (ii) fast similarity-based search operations will be enabled, so that previous tasks with similar input data can be found and reused {\revision (no execution from scratch will be needed)}. In the example of Figure~\ref{fig:arch}, snapshots $n-1$ and $n$ may have the same hash values given that they are highly similar. A minimum similarity threshold {\revision may be selected by each application,} so that a task $t$ can reuse a task $t_{reused}$ only if the similarity between the input data of $t$ and $t_{reused}$ is higher {\revision than this threshold.} {\rone This offers flexibility and enables different applications to set different similarity thresholds that may be acceptable based on their requirements. Different forms of similarity can also be applied between $t$ and $t_{reused}$, such as structural similarity or cosine similarity.}

%{\rone The appropriate similarity threshold to be selected among tasks depends on the nature of each application and the granularity of the needed input data processing. For example, in the case of an application that requires coarse-grain processing of input data (\eg identification of whether vehicle traffic exists in a captured snapshot), a lower similarity threshold may be appropriate. On the other hand, in the case of an application that requires fine-grained data processing (\eg identification of the exact location of vehicles in captured snapshots), higher similarity thresholds may be appropriate. The similarity threshold for each application can be selected by different stakeholders. For instance, application developers, being aware of the characteristics of their applications and the granularity of the processing that input data may need, can specify the appropriate threshold among tasks. In addition, the administrators of an edge computing system/environment typically keep extensive statistics about the operation of this system/environment. As part of these statistics, the similarity among offloaded tasks for various edge computing applications may be also maintained. }

Depending on the available resources, devices can cache offloaded tasks and the results of their execution (once received by edge servers). This can act as a first layer of {\revision computation reuse before new tasks are offloaded to the edge.} {\revision For example, multiple applications may run on an AR headset, {\rone requiring the detection of objects} in scenes captured by the headset (\eg a driving assistance application that identifies potential accident conditions and informs drivers, and a smart navigation application that provides instructions to drivers on how to reach their final destinations)}~\cite{guo2018potluck}. Such applications can essentially share and reuse the results of the tasks they offload. If deduplication and reuse are not possible at the device level {\revision (no similar previous tasks were found or no resources are available on devices to store previously executed tasks and their results)}, a task along with {\revision the hash of} its input {\rone data and the desired similarity threshold are} sent towards the edge network infrastructure. {\revision If a user device does not have adequate computing power to produce a hash of the task input data, the first edge router that receives the task can generate the hash. This router will attach the generated hash to the task, so that edge routers and servers that subsequently receive this task do not need to generate the hash again.}
%To achieve that, user devices cache executed tasks and their results 

\subsection {Layer 2: Edge Network Infrastructure}
\label{subsec:network}

The primary goal of the edge network infrastructure is to forward tasks for the same service (or services with common components) and with similar input data to the edge server {\rone (among the available edge servers)} that can maximize the chances of reusing previous computation. At the same time, in-network storage resources may be available to cache/store previously executed tasks and their execution results as they {\rone transit through the edge network infrastructure.} When an offloaded task is received by a router, the router may search for previously executed similar tasks, if local storage resources are available. This similarity search process will take place based on the locality sensitive or feature hash that has been attached by user devices to the offloaded task. If no previously executed task that can be reused is found, a router will forward a task {\revision based on its hash to an edge server.} The space of the potential hash values is divided among the available edge servers, so that each edge server is responsible for the execution of tasks with input data that falls under the range of the hash values assigned to this server. For example, in Figure~\ref{fig:arch}, each hash has a size of 4 bytes. {\rone To this end, the potential hash values} will be between 0 and 65,535, while these values are equally divided among the available edge servers.
%and edge servers A, B, and C will be responsible for hash values 0-21,845, 21,846-43,690, and 43,691-65,535 respectively. 

\subsection {Layer 3: Edge Servers}

Edge servers receive offloaded tasks and perform {\revision a nearest neighbor search} to identify previously executed stored tasks that could be reused. {\revision Once the nearest} neighbor of an incoming task $t$ is found, an edge server will check whether the similarity between the input data of $t$ and the nearest neighbor of $t$ exceeds the minimum similarity {\rone threshold selected by the application} that offloaded $t$. If this is the case, the found nearest neighbor task will be reused and its results will {\revision be returned to the user} in response to $t$. Otherwise, the server will execute $t$ and store $t$ and its execution results for potential reuse in the future. Each edge server maintains one or more hash {\rone tables, which index previously executed stored tasks based on the hashes of the} tasks' input data. Overall, the edge servers trade storage for computing {\revision to reduce} response times for users and increase the number of users, devices, and tasks that can be simultaneously accommodated. %by the available edge servers.

\subsection {Practicality Check: Could Such An Architecture Work?}
\label{subsec:results}

We implemented and evaluated such an architecture {\rone based on a topology that consists of two user devices, two edge routers, and two edge servers. Each device and edge router is equipped with an Intel Core i5-4250U CPU @1.30GHz and 8GB of memory, while each edge server is equipped with an Intel Core i5-9600K CPU @3.70GHz and 64GB of memory. Each user device is connected to an edge router, while each edge router is connected to both edge servers.} Each user device offloads tasks that are received by an edge router and are forwarded to one of the edge servers (each router has connections to both servers). The Round Trip Time (RTT) between devices and servers ranges between 12-16 ms. {\rone We have created Application Programming Interfaces (APIs) to realize the LSH semantics (hashing and nearest neighbor search)} using the FALCONN library as our basis~\cite{andoni2015practical}. The edge servers run tensorflow machine learning models offering an image annotation service, while images are offloaded from user devices to edge servers. Routers forward tasks towards edge servers as determined in Section~\ref{subsec:network} {\revision and store previously} executed tasks directly in the network. We used the following {\revision image datasets as the input data of tasks}:

%\begin{figure}[th]
%    \centering
%    \includegraphics[width=1\linewidth]{fig/topo.pdf}
%    \caption{\small Experimental topology.}
%    \label{fig:topo}
%    \vspace{-0.2cm}
%\end{figure}

\begin {itemize}

\item The {\rone Modified National Institute of Standards and Technology (MNIST)} dataset {\revision consisting} of 70K images of handwritten digits~\cite{lecun1998mnist}.

\item The Pandaset dataset {\revision consisting} of 48K images taken from cameras on-board autonomous vehicles in California, USA~\cite{pandaset}. 

\item A dataset of mobile AR consisting of 1K object images published by Stanford University~\cite{makar2013interframe}. 

\item A dataset of 10K snapshots of vehicle traffic that we captured through CCTV cameras {\revision monitoring traffic in Omaha, Nebraska, USA.}

\end {itemize}

{\rone Our evaluation results indicate that a low-end computer (equipped with a dual core processor and 8GBs of RAM) can generate a locality sensitive hash for an image in less than 1.8ms. A nearest neighbor search can also be performed in less than 1ms for up to 100K stored images once a locality sensitive hash has been generated. We have also quantified the overhead of storing a task and its execution results for different edge services (an object detection service, a voice command service for the control of smart home IoT devices, and a 3D graphics rendering service). Our results demonstrate that 0.0023-0.06MB of storage space is needed per task depending on the type of the service, and the size of the input data and execution results. Techniques, such as compression and downsampling, can be applied to achieve the low end of the presented range.}
%In the case of: (i) an object detection edge service, 0.0023-0.0478MB of storage space is needed per task; (ii) a voice command service for the control of smart home IoT devices, 0.0025-0.0479MB of storage space is needed per task; and (iii) a 3D graphics rendering service, 0.014-0.06MB of storage space is needed per task. The exact storage space value per task depends on the size of the input data and execution results, while compression and downsampling techniques can be applied to achieve the low ends of the presented ranges.} 
In Table~\ref{fig:results}, we present the completion time of offloaded tasks (\ie the time elapsed between the generation of a task by a device and the {\revision retrieval of the task's execution results by this device) for all datasets.} Our results indicate that on average the deduplication and reuse of computation resulted in 5.67-16.05$\times$ lower task completion times than cases where computation deduplication and reuse are not applied. 

\begin{figure}[th]
    \centering
    \includegraphics[width=0.9\linewidth]{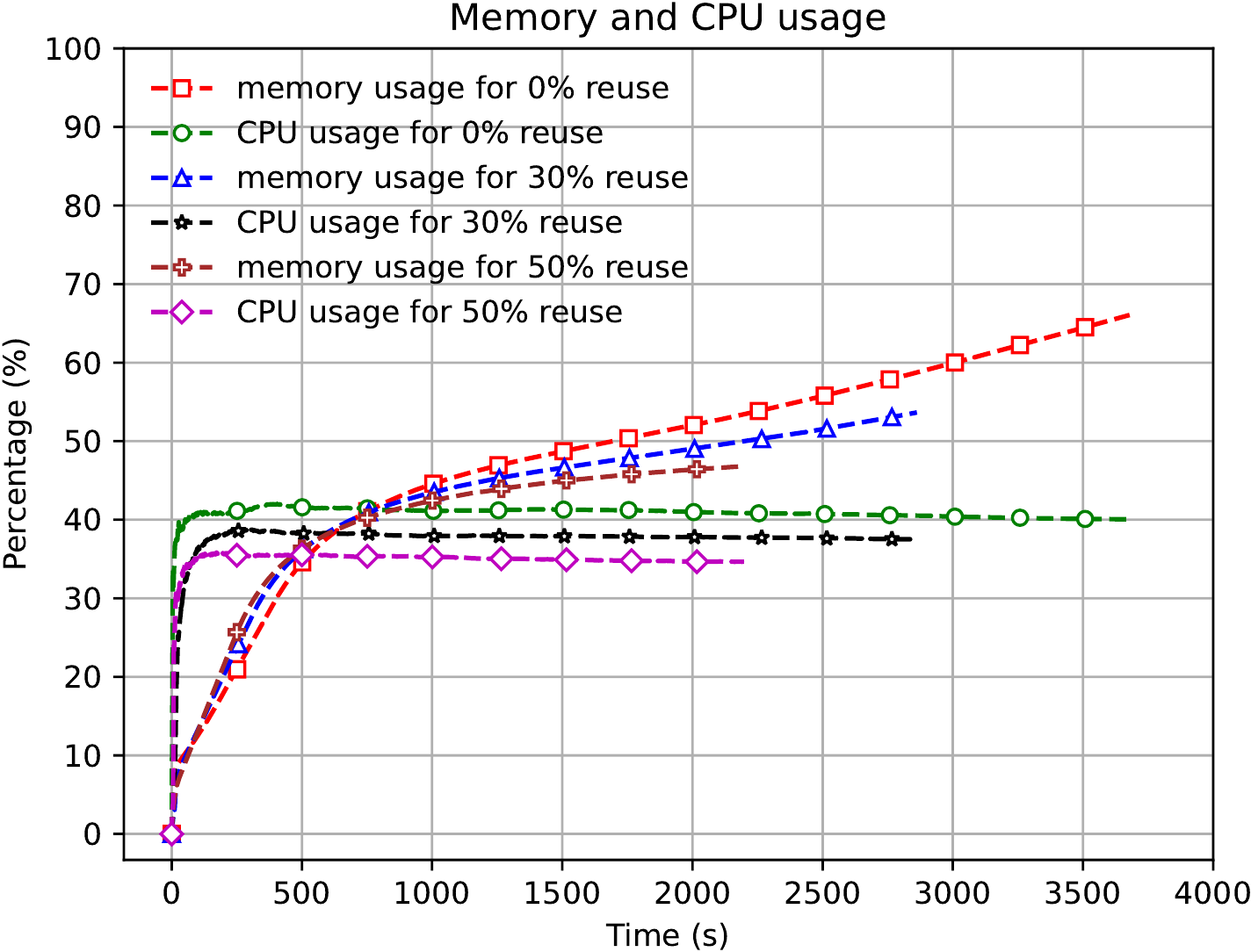}
    \caption{{  Percentage of CPU and memory usage of an edge server during the execution of 40K offloaded tasks (markers do not represent actual data points, but are used for better readability).}}
    \label{fig:usage}
    \vspace{-0.2cm}
\end{figure}

{\rone As we present in Table~\ref{table:resultsmore}, the percentage of tasks that can be reused relies on the similarity between the input data of tasks and decreases as we increase the minimum similarity threshold.} Our results show that the proposed architecture can accommodate tasks with input data that exhibits high degrees of similarity (\eg CCTV dataset), moderate degrees of similarity (\eg Mobile AR dataset), as well as lower degrees of similarity (\eg Pandaset and MNIST datasets). {\rone The reuse accuracy (\ie the percentage of offloaded tasks, which reused tasks with results that were the same as the results that their own execution would have produced)} {\revision improves as we increase the minimum similarity threshold}. The reuse accuracy finally reaches 95-100\% among all datasets for a similarity threshold of 90\%. {\rone Finally, in Figure~\ref{fig:usage}, we present the usage of the CPU and memory resources of an edge server during the execution of 40K offloaded tasks. The results demonstrate that the usage percentage drops as the percentage of reuse increases. The resources of the edge server are also occupied for smaller amounts of time as the reuse percentage increases, since the execution of tasks is completed sooner as compared to cases without reuse.}

\begin{table}[h]
\centering
\caption{\small {  Task completion times when reuse occurs: (i) at user devices and within the edge network infrastructure; and (ii) at edge servers.}}
\label{fig:results}
\begin{tabular}{|c|c|c|c|}
\hline
\multicolumn{4}{|c|}{\color{black}{\textbf{Average Task Completion Time (ms)}}           }                                                                                                                                    \\ \hline
{\color{black}\textbf{Dataset}} & 
{\color{black}\textbf{No reuse}} & {\color{black}\textbf{\begin{tabular}[c]{@{}c@{}}Reuse (Edge \\ Servers)\end{tabular}}} & {\color{black}\textbf{\begin{tabular}[c]{@{}c@{}}Reuse (Devices and \\ Edge Network)\end{tabular}}} \\ \hline
 {\color{black} MNIST}            &  {\color{black} 120.42}            &  {\color{black} 21.23}                                                                    &  {\color{black} 8.82}                                                                                 \\ \hline
 {\color{black} Pandaset}         &  {\color{black} 116.65}            &  {\color{black} 19.52}                                                                    &  {\color{black} 7.26}                                                                                 \\ \hline
 {\color{black} Mobile AR}        &  {\color{black} 106.64}            &  {\color{black} 18.32}                                                                    &  {\color{black} 6.68}                                                                                 \\ \hline
 {\color{black} CCTV}     &  {\color{black} 115.68}            &  {\color{black} 17.80}                                                                    &  {\color{black} 7.67}                                                                                 \\ \hline
\end{tabular}
\end{table}

\begin{table*}[t]
%\resizebox{\textwidth}{!}{
\centering
\caption{{  Percentage of reused tasks and accuracy of reuse for all datasets and varying similarity thresholds.}}
\label{table:resultsmore}
\begin{tabular}{|c|cccc|cccc|}
\hline
\textbf{}                                                                     & \multicolumn{4}{c|}{\textbf{Percentage of reuse (\%)}}                                              & \multicolumn{4}{c|}{\textbf{Reuse accuracy (\%)}}                                                   \\ \hline
\textbf{\begin{tabular}[c]{@{}c@{}}Similarity \\ threshold (\%)\end{tabular}} & \multicolumn{1}{c|}{MNIST} & \multicolumn{1}{c|}{Pandaset} & \multicolumn{1}{c|}{Mobile AR} & CCTV  & \multicolumn{1}{c|}{MNIST} & \multicolumn{1}{c|}{Pandaset} & \multicolumn{1}{c|}{Mobile AR} & CCTV  \\ \hline
60                                                                            & \multicolumn{1}{c|}{45.54} & \multicolumn{1}{c|}{29.06}    & \multicolumn{1}{c|}{33.43}     & 91.17 & \multicolumn{1}{c|}{84.44} & \multicolumn{1}{c|}{72.57}    & \multicolumn{1}{c|}{95.43}     & 84.37 \\ \hline
70                                                                            & \multicolumn{1}{c|}{41.11} & \multicolumn{1}{c|}{24.44}    & \multicolumn{1}{c|}{31.36}     & 90.08 & \multicolumn{1}{c|}{88.76} & \multicolumn{1}{c|}{77.05}    & \multicolumn{1}{c|}{98.92}     & 86.37 \\ \hline
80                                                                            & \multicolumn{1}{c|}{35.4}  & \multicolumn{1}{c|}{20.78}    & \multicolumn{1}{c|}{28.41}     & 87.54 & \multicolumn{1}{c|}{90.60} & \multicolumn{1}{c|}{86.10}    & \multicolumn{1}{c|}{100}       & 89.41 \\ \hline
90                                                                            & \multicolumn{1}{c|}{33.58} & \multicolumn{1}{c|}{20.17}    & \multicolumn{1}{c|}{24.25}     & 80.76 & \multicolumn{1}{c|}{95.01} & \multicolumn{1}{c|}{95.03}    & \multicolumn{1}{c|}{100}       & 96.91 \\ \hline
\end{tabular}
%}
\end{table*}

%\begin{figure}[th]
%	\centering
%	%\vspace{-0.35cm}
%	\captionsetup[subfigure]{aboveskip=-1pt,belowskip=-1pt}
%	\begin{subfigure}[t]{1\columnwidth}
%		\centering
%		\includegraphics[width=1\columnwidth]{fig/percent-reuse-magazine-crop.pdf}
%		\vspace{+0.1cm}
%		\subcaption{{  Percentage of reused (not executed from scratch) tasks.}}
%		\vspace{+0.1cm}
%		\label{Figure:percent}
%	\end{subfigure}
%	\begin{subfigure}[b]{1\columnwidth}
%		\centering
%		\includegraphics[width=1\columnwidth]{fig/accuracy-magazine-crop.pdf}
%		\vspace{+0.1cm}
%		\subcaption{{  Accuracy of reuse.}}
%		\vspace{+0.1cm}
%		\label{Figure:accuracy}
%	\end{subfigure}
	%\vspace{-0.2cm}
%	\caption{{  Percentage of reused tasks and accuracy of reuse for all datasets and varying similarity thresholds.}}
%	\label{figure:moreresults}
%	%\vspace{-0.35cm}
%\end{figure}

\subsection {{\rone Shortcomings and Limitations}}
%\label{subsec:results}

{\rone In its current form, our proposed architecture could cause load imbalances among edge servers, since large amounts of similar tasks may be forwarded to the same edge server(s), thus increasing the load of certain servers and leaving others under-utilized. In addition, mechanisms are needed to dynamically distribute the hash value space among edge servers. Our architecture could also benefit from techniques to optimize the usage of storage resources by storing tasks that are likely to be reused in the future, while discarding ones that are not likely to be reused. Finally, the reuse of computation could be exploited by attackers to discover if tasks with similar data and/or for the same edge service have been previously executed. We further discuss these open issues and propose possible solutions in Section~\ref{sec:discussion}. }

\section{Open Challenges and Future Directions}
\label{sec:discussion}

{\rone Computation deduplication and reuse show} promise for edge computing environments, {\revision having the potential to improve response times. However, there are still open challenges to be} addressed leading to several directions of future research.

\noindent \textbf {The ``curse'' of dimensionality:} Through hashing, {\revision high-dimensional data is} converted to a fixed-size value. This process may require a large space of features for FH and a family of hash functions for LSH to be applied to {\rone the} task input data {\rone in order to maintain} satisfactory reuse accuracy. Large feature spaces and {\revision LSH function} families may result in longer hashing and search times {\rone and} increase the memory requirements for user devices, edge routers, and servers. {\revision Mechanisms, such as hierarchical feature hashing~\cite{zhao2014hierarchical} and multi-probe LSH~\cite{lv2007multi}, to keep the size of the feature space and the number of {\revision LSH functions} manageable should be further explored.}

\noindent \textbf{Distribution of hash value space among edge servers:} The {\rone hash value space} needs to be divided and distributed among the available edge servers. {\rone To achieve that, mechanisms of different nature (distributed and centralized) should be explored.} Distributed mechanisms can enable servers to essentially form a multicast communication group. In the context of this group, servers communicate directly to reach a consensus on how to divide the hash value space and which server will be responsible for which range of the hash value space. Logically centralized mechanisms may utilize Software-Defined Networking (SDN) controllers, which {\revision act as coordination points for the distribution of the hash value space among servers. SDN controllers can inform edge routers about the distribution of the hash value space among servers, populating the reuse information table of routers. Initially, the hash value space can be equally distributed {\rone among servers and} it can be dynamically redistributed to balance the load among servers as we describe below.}

%At the same time, SDN controllers can monitor the computation deduplication and reuse performance and overhead among edge servers over time 

\noindent \textbf {Balancing the load {\rone among} edge servers:} {\rone As the space of potential hash values} is divided among edge servers, load imbalances may occur. For example, if large amounts of tasks with similar input data are generated, all the tasks may be forwarded to the {\revision same server(s)}, increasing the load of certain servers, while leaving other servers under-utilized. This calls for mechanisms to achieve load balancing and {\revision reuse at the same time. For example, SDN controllers can monitor the computation reuse performance,} {\rone the overhead among servers, and the load of servers,} and redistribute parts of the hash value space from one server to another to balance the load. 

\noindent \textbf {Predicting the likelihood of reuse:} The storage resources of {\revision user devices, edge routers, and servers} may have a limited capacity. {\revision In environments that offer computation reuse, the different layers of the architecture} may not be able to store the results of all executed tasks. To increase the impact and benefits of reuse, mechanisms to estimate/predict the chances {\rone of an executed task being reused in the future} need {\rone to be explored}. As a result, tasks not likely to be reused in the future may not be stored after execution, offering available storage space to tasks, which are likely to be reused. {\rone Such mechanisms may also be} essential in cases of tasks that consist of multiple sub-tasks (\eg tasks that are formulated as a computation graph) to determine which sub-tasks to store and which ones to discard across the different layers of the architecture.

%{\revision \noindent \textbf{Extending the notion of computation reuse:} In this paper, we have focused on the reuse of computation for tasks with similar input data that yield the same execution results. This ensures that the operation of applications and the user-perceived quality of experience are not negatively impacted, since {\rone previous tasks are} reused only if they yield the same results as offloaded tasks. However, research directions, such as approximate computing, %~\cite{han2013approximate}, 
%suggest that certain applications may be able to operate with approximate (not fully precise) processing results of the data they generate without significant degradation of user experience. Such directions may enable the expansion of the notion of computation reuse as defined in our work, so {\rone that tasks which yield} similar results can also be reused. A thorough assessment of the reuse of tasks that yield similar versus the same results and its impact on the perceived user quality of {\rone experience should be investigated.}}

%\noindent \textbf {Cache/storage optimizations:} As we mentioned above, edge servers may not be able to store all the executed tasks and their results. In addition to predicting how likely it is for a task and its results to be reused in the future, effective cache/storage eviction and replacement policies need to be explored. These policies will aid decisions about which tasks to replace as tasks are executed by servers, occupying the full capacity of the servers' storage resources.

\noindent \textbf {\rone Security and privacy implications}: {\revision Attackers %, being aware of the fact that computation reuse mechanisms may be available across multiple layers of the edge architecture, 
can probe the edge architecture} {\rone to discover if tasks for the same service and/or with similar input data} have been previously executed. For example, attackers can offload tasks {\revision with images to be processed by an object detection edge service}, while knowing that such tasks may need {\rone several tens or even hundreds} of milliseconds to be executed. As a result, if the execution results are received much sooner, attackers can infer that a task with similar {\revision input data was reused.} {\rone In addition, given that the execution results of tasks offloaded by different users can be shared/reused, solutions to isolate private results but share non-private results in multi-tenant (multi-user) edge environments should be explored~\cite{chen2019enclavecache}. Attackers could also infer the locations of the devices that offload tasks~\cite{tian2020location}. The implications of reuse on the security and privacy of computations, the associated input data, and the location of devices should be further investigated.}

{\revision \noindent \textbf{Scalability:} Given the projected growth {\rone of devices and the} wide spectrum of next-generation applications, scalability becomes a major challenge for computation deduplication and reuse architectures. {\rone Techniques to optimize} the performance of hashing and nearest neighbor search operations can contribute to scaling up the number of tasks that can be handled. The scalability of computation reuse architectures can be further enhanced by performing reuse not only on the basis of individual applications, but for groups of applications that require the same type of data processing. For example, {\rone applications that need the} detection of objects in images may invoke different services of the same type/nature deployed at the edge. Such services essentially provide the same type of data processing (\ie detection of objects in images), however, they may achieve that through different object detection algorithms.} 

\section{Conclusion}
\label{sec:conclusion}

In this paper, we presented the promise and challenges of computation deduplication and reuse in {\rone edge computing deployments}. We first {\revision presented use-cases}, which computation reuse {\revision can benefit,} and we then discussed the technical challenges of {\revision realizing solutions} {\rone for the reuse of} computation. Moreover, we presented the design of a multi-layer architecture for computation reuse and several open challenges {\revision and research directions}. We believe that the effective management of the massive computation volumes projected to be produced at the edge will become {\revision a pressing issue}, {\rone thus reusing computation} among devices, users, and applications will become a key mechanism to improve response times and accommodate {\revision additional users, devices, and tasks.}

\section*{Acknowledgements}
This work is partially supported by NIH (NIGMS/P20GM109090), NSF under awards CNS-2016714 and CNS-2104700, the Nebraska University Collaboration Initiative, and the Nebraska Tobacco Settlement Biomedical Research Development Funds.

\bibliographystyle{IEEEtran}
\bibliography{Ref}

\begin{thebibliography}{10}
\providecommand{\url}[1]{#1}
\csname url@samestyle\endcsname
\providecommand{\newblock}{\relax}
\providecommand{\bibinfo}[2]{#2}
\providecommand{\BIBentrySTDinterwordspacing}{\spaceskip=0pt\relax}
\providecommand{\BIBentryALTinterwordstretchfactor}{4}
\providecommand{\BIBentryALTinterwordspacing}{\spaceskip=\fontdimen2\font plus
\BIBentryALTinterwordstretchfactor\fontdimen3\font minus
  \fontdimen4\font\relax}
\providecommand{\BIBforeignlanguage}[2]{{%
\expandafter\ifx\csname l@#1\endcsname\relax
\typeout{** WARNING: IEEEtran.bst: No hyphenation pattern has been}%
\typeout{** loaded for the language `#1'. Using the pattern for}%
\typeout{** the default language instead.}%
\else
\language=\csname l@#1\endcsname
\fi
#2}}
\providecommand{\BIBdecl}{\relax}
\BIBdecl

\bibitem{shi2016edge}
W.~Shi \emph{et~al.}, ``Edge computing: Vision and challenges,'' \emph{IEEE
  internet of things journal}, vol.~3, no.~5, pp. 637--646, 2016.

\bibitem{satyanarayanan2017emergence}
M.~Satyanarayanan, ``The emergence of edge computing,'' \emph{Computer},
  vol.~50, no.~1, pp. 30--39, 2017.

\bibitem{guo2018foggycache}
P.~Guo \emph{et~al.}, ``Foggycache: Cross-device approximate computation
  reuse,'' in \emph{Proceedings of the 24th Annual International Conference on
  Mobile Computing and Networking}, 2018, pp. 19--34.

\bibitem{guo2018potluck}
P.~Guo and W.~Hu, ``Potluck: Cross-application approximate deduplication for
  computation-intensive mobile applications,'' in \emph{Proceedings of the
  Twenty-Third International Conference on Architectural Support for
  Programming Languages and Operating Systems}, 2018, pp. 271--284.

\bibitem{drolia2017cachier}
U.~Drolia \emph{et~al.}, ``Cachier: Edge-caching for recognition
  applications,'' in \emph{2017 IEEE 37th international conference on
  distributed computing systems (ICDCS)}.\hskip 1em plus 0.5em minus
  0.4em\relax IEEE, 2017, pp. 276--286.

\bibitem{meng2020coterie}
J.~Meng \emph{et~al.}, ``Coterie: Exploiting frame similarity to enable
  high-quality multiplayer vr on commodity mobile devices,'' in
  \emph{Proceedings of the Twenty-Fifth International Conference on
  Architectural Support for Programming Languages and Operating Systems}, 2020,
  pp. 923--937.

\bibitem{mastorakis2020icedge}
S.~Mastorakis \emph{et~al.}, ``{ICedge:} when edge computing meets
  information-centric networking,'' \emph{IEEE Internet of Things Journal},
  vol.~7, no.~5, pp. 4203--4217, 2020.

\bibitem{andoni2015practical}
A.~Andoni \emph{et~al.}, ``Practical and optimal lsh for angular distance,''
  \emph{arXiv preprint arXiv:1509.02897}, 2015.

\bibitem{lecun1998mnist}
Y.~LeCun, ``The mnist database of handwritten digits,''
  \emph{http://yann.lecun.com/exdb/mnist/, Accessed on March 10, 2021}, 1998.

\bibitem{pandaset}
``Pandaset by hesai and scale ai,'' https://pandaset.org, Accessed on March 10,
  2021.

\bibitem{makar2013interframe}
M.~Makar \emph{et~al.}, ``Interframe coding of canonical patches for low
  bit-rate mobile augmented reality,'' \emph{International Journal of Semantic
  Computing}, vol.~7, pp. 5--24, 2013.

\bibitem{zhao2014hierarchical}
B.~Zhao \emph{et~al.}, ``Hierarchical feature hashing for fast dimensionality
  reduction,'' in \emph{Proceedings of the IEEE Conference on Computer Vision
  and Pattern Recognition}, 2014, pp. 2043--2050.

\bibitem{lv2007multi}
Q.~Lv \emph{et~al.}, ``Multi-probe lsh: efficient indexing for high-dimensional
  similarity search,'' in \emph{33rd International Conference on Very Large
  Data Bases, VLDB 2007}.\hskip 1em plus 0.5em minus 0.4em\relax Association
  for Computing Machinery, Inc, 2007, pp. 950--961.

\bibitem{chen2019enclavecache}
L.~Chen, J.~Li, R.~Ma, H.~Guan, and H.-A. Jacobsen, ``Enclavecache: A secure
  and scalable key-value cache in multi-tenant clouds using intel sgx,'' in
  \emph{Proceedings of the 20th International Middleware Conference}, 2019, pp.
  14--27.

\bibitem{tian2020location}
Z.~Tian, Y.~Wang, Y.~Sun, and J.~Qiu, ``Location privacy challenges in mobile
  edge computing: classification and exploration,'' \emph{IEEE Network},
  vol.~34, no.~2, pp. 52--56, 2020.

\end{thebibliography}

%\vspace{-1cm}
\section*{Biographies}

\vskip -2.0\baselineskip plus -1fil

\begin{IEEEbiographynophoto}{Md Washik Al Azad}
(malazad@unomaha.edu) is a Ph.D. student at the University of Nebraska Omaha. He received his B.S. in Electronics and Telecommunication Engineering from the Rajshahi University of Engineering \& Technology, Bangladesh in 2017. His interests include edge computing and security.
\end{IEEEbiographynophoto}

\vskip -2.0\baselineskip plus -1fil

\begin{IEEEbiographynophoto}{Spyridon Mastorakis}
(smastorakis@unomaha.edu) is an Assistant Professor in Computer Science at the University of Nebraska Omaha. He received his Ph.D. in Computer Science from the University of California, Los Angeles in 2019.
His research interests include network systems and architectures, edge computing, and security.
\end{IEEEbiographynophoto}

\end{document}